\documentclass[aps,pre,twocolumn,float]{revtex4}
\usepackage{amsmath,bm,epsfig}

\usepackage{pstricks}
\usepackage{pst-node}
\usepackage[ansinew]{inputenc}
\usepackage{amssymb,amsmath}
\usepackage{multirow}

\def\be{\begin{equation}}       \def\ba{\begin{array}}

\def\ee{\end{equation}}         \def\ea{\end{array}}

\def\bea {\begin{eqnarray}}      \def\eea {\end{eqnarray}}

\def\bean{\begin{eqnarray*}}    \def\eean{\end{eqnarray*}}

\newtheorem{exi}{Example}

\DeclareMathOperator\arctanh{arctanh}

\begin{document}

\title{Hamiltonian systems with an infinite number of localized travelling waves}
\author{Georgy L. Alfimov$^{a}$, Elina V. Medvedeva$^{a}$, and Dmitry E. Pelinovsky$^{b,c}$}
    \affiliation{$^a$ Moscow Institute of Electronic Engineering,
    Zelenograd, Moscow, 124498, Russia}
    \affiliation{$^b$ Department of Mathematics, McMaster University, Hamilton, Ontario, Canada}
    \affiliation{$^c$ Department of Applied Mathematics,  Nizhny Novgorod State Technical University, Nizhny Novgorod, Russia}

\begin{abstract}
In many Hamiltonian systems, propagation of steadily travelling
solitons or kinks is prohibited because of resonances with linear
excitations. We show that Hamiltonian systems with resonances may admit
an infinite number of travelling solitons or kinks if the closest to
the real axis singularities in the complex upper half-plane of limiting
asymptotic solution are of the form $z_\pm=\pm\alpha+i\beta$,
$\alpha\ne 0$. This quite a general statement is illustrated by
examples of the fifth-order Korteweg--de Vries-type equation, the
discrete cubic-quintic Klein--Gordon equation, and the nonlocal double
sine--Gordon equations.
\end{abstract}

\maketitle

{\bf Introduction.} Nonlinear localized travelling waves such as bright
or dark solitons are key concepts for many branches of modern physics,
including nonlinear optics, theory of magnets, theory of Josephson
junctions, etc. It is known that in many dispersive systems the
presence of these nonlinear entities is strongly restricted due to {\it
resonances} with linear excitations. These resonances take place in
Hamiltonian systems of various origin, such as the fifth-order
Korteweg--de Vries equation \cite{Kaw1}, nonlinear lattices
\cite{Savin,Savin2000,IP} and models with complex dispersion and
nonlocal interactions \cite{A1,AOSU95}. As a result, it is quite
typical that in such Hamiltonian systems the localized excitations
either do not exist at all or they only exist for specific values of
some external parameters. In the last case the nonlinear excitations
are called {\it embedded solitons} (i.e. solitons ``embedded'' into the
spectrum of linear waves). These embedded solitons have been discovered
in hydrodynamics, nonlinear optics and other fields of modern physics
\cite{ES3Gen}.

To give an example, consider an operator equation
\begin{eqnarray}
{L}_{\varepsilon} u = F(u), \label{MainEq}
\end{eqnarray}
for a function $u(\xi)$ where ${L}_{\varepsilon}$ is a Fourier
multiplier operator in $\xi$ space with even symbol $\widehat{L}(k)$ in
$k$ space, $F(u)$ is a nonlinear function and $\varepsilon$ is a
parameter. The prototypical examples of problems leading to
Eq.(\ref{MainEq}) are the generalized Korteweg--de Vries equation,
\begin{eqnarray}
u_t+(F(u))_x+{M}_\varepsilon u_{x}=0,\label{Exam1}
\end{eqnarray}
or discrete or nonlocal Klein-Gordon equations,
\begin{eqnarray}
u_{tt}-{M}_\varepsilon u+F(u)=0, \label{Exam2}
\end{eqnarray}
for $u(x,t)$, where ${M}_\varepsilon$ is a Fourier multiplier operator
in $x$ space. Above $\xi=x-vt$ is the travelling wave coordinate and
the operator $L_\varepsilon$ in Eq.(\ref{MainEq}) includes both
$M_\varepsilon$ and $v$. We assume that in the both cases
$\varepsilon=0$ implies a degeneration of the problem with $L_0 =
\partial_\xi^2$.

Consider a solitary wave $u(\xi)$ which is asymptotic to the
equilibrium state $u\equiv 0$ as $\xi \to\pm\infty$ (the case of kink
wave which is asymptotic to a pair of equilibrium states $u \equiv
u_{\pm}$ as $\xi \to \pm \infty$ can be analyzed in a similar way).
Then the resonances correspond to the real roots of the dispersion
equation near $u\equiv 0$
\begin{eqnarray}
\widehat{L}_{\varepsilon}(k) = F'(0).
\label{DisRel}
\end{eqnarray}
If for some value of $\varepsilon$, there exist a {\it single} pair of
real roots $k= \pm k_0$ in Eq.(\ref{DisRel}), we are in situation when
the resonance prohibits propagation of regular solitons in
Eq.(\ref{Exam1}) and Eq.(\ref{Exam2}) and the embedded solitons may
appear. In this case, the velocity $v$ of the soliton, typically, {\it
is not arbitrary} but should be ``adjusted'' to avoid ``gluing'' with
linear modes. In general, $v$ belongs to some discrete set. This set
may be empty (i.e, no localized waves propagate), or include finite or
infinite number of values. The case when Eq.(\ref{DisRel}) has more
then one pair of real roots is more complex and the presence of
localized excitations in this case is highly doubtful.

In this paper we address the following question: {\it Are there some
conditions which would indicate to existence of infinitely many
embedded solitons described by Eq.(\ref{MainEq})? If this is possible,
can we describe this infinite set asymptotically?} As a result, we
present conditions for existence of countable infinite sequence of
embedded solitons. The main assumption is that the limiting solution of
Eq.(\ref{MainEq}) as $\varepsilon \to 0$, being extended in a complex
plane, should have a pair of symmetric singularities in the upper
half-plane. We give an asymptotic formula for values $\{ \varepsilon_n
\}$ as $n \to \infty$, for which embedded solitons exist. In terms of
Eq.(\ref{Exam1}) and Eq.(\ref{Exam2}) this means a presence of an infinite number
of velocities $v$ for nonlinear localized excitations.

Up to the moment no rigorous proof of this asymptotic theory has been
found. We give some heuristic explanation of the mechanism behind the
asymptotic formula and illustrate this result with three numerical
examples. Surprisingly, it has been observed that in some cases the
asymptotic formula predicts the parameters even of lowest embedded
solitons from this sequence  with reasonable accuracy.

We note that the idea that two symmetric singularities in the upper
half-plane can be related to the countable infinite sequence of
tangential intersections of stable and unstable manifolds can be found
in \cite{G1} for the primary intersection point of the two-dimensional
symplectic maps. In this paper we generalize this principle to more
general class of physically relevant systems. \vspace{0.5cm}

{\bf Main result. } Consider Eq. (\ref{MainEq}), where $u(\xi)$ is
real-valued function defined on $\mathbb{R}$. Assume that
$L_\varepsilon$ is a real operator which depends continuously on real
parameter $\varepsilon$ and satisfies $L_0 = \partial_\xi^2$. The
Fourier symbol ${\hat L}_\varepsilon(k)$ is supposed to be an even
function of $k$.

Assume now that (a) the equation $F(u)=0$ has zero solution $u=0$
with $F'(0) > 0$; (b) the dispersion equation (\ref{DisRel})
has only one pair of real roots $k= \pm k(\varepsilon)$ such that
$k(\varepsilon)\to\infty$ as $\varepsilon\to 0$; and (c) the equation
\begin{eqnarray}
\label{lim-eq}
u'' = F(u)
\end{eqnarray}
has an even localized solution ${\tilde u}(\xi)$ such that ${\tilde
u}(\xi)\to 0$ as $\xi\to\pm\infty$.  In addition, the key assumption of
our asymptotic theory is that the solution $\tilde{u}(\xi)$ can be
continued into the complex plane and the closest to the real axis
singularities of ${\tilde u}(\xi)$ in the upper half-plane are given by
the pair $z_{\pm} = \pm \alpha + i \beta$, with $\alpha, \beta > 0$,
which is symmetric with respect to the imaginary axis.

{\it Then}, we expect the existence of an infinite sequence of values
$\{ \varepsilon_n \}$ such that for each $\varepsilon = \varepsilon_n$,
Eq.(\ref{MainEq}) has a soliton solution $u(\xi)$ with $u(\xi)\to 0$ as
$\xi\to\pm \infty$, and this sequence obeys the following asymptotic
law
\begin{eqnarray}
k(\varepsilon_n)\sim \left(n\pi+\varphi_0\right)/\alpha \label{AsRel}
\end{eqnarray}
where $\varphi_0$ is a phase constant that depends on $L_\varepsilon$ and $\tilde u$.

This result can be extended naturally to the case of kink solutions of
Eq.(\ref{MainEq}) connecting a pair of equilibrium states $u=u_{\pm}$,
such that $F(u_{\pm})=0$ and $F'(u_-)=F'(u_+)>0$. In this case,
$F'(u_{\pm})$ appears instead of $F'(0)$ in the dispersion relation
(\ref{DisRel}), whereas the differential equation (\ref{lim-eq}) is
assumed to have a kink solution $\tilde{u}(\xi)$ such that ${\tilde
u}(\xi)\to u_{\pm}$ as $\xi\to\pm\infty$.

\vspace{0.5cm}

{\bf Justification}. Let us give some heuristic arguments for
justification of the main result. Introduce $v(\xi)=u(\xi)-{\tilde
u}(\xi)$, $N(v)=F(\tilde u+v)-F(\tilde u)-F'(\tilde u)v$ and
$H_\varepsilon=L_0-L_\varepsilon$. Then, we have
\begin{eqnarray}
\label{perturbation}
(L_\varepsilon-F'(\tilde u))v=H_\varepsilon\tilde u+N(v).
\end{eqnarray}
Based on our assumptions, we take as granted that (a) the function
$H_\varepsilon\tilde u(\xi)$ can be continued into the upper complex
half-plane and its closest to the real axis singularities are
$z_\pm=\pm\alpha+i\beta$, whereas (b) the homogeneous linearized
equation
\begin{eqnarray}
(L_\varepsilon-F'(\tilde u))v=0\label{LHS}
\end{eqnarray}
has  a pair of solutions $\varphi_\varepsilon^\pm(\xi)=e^{\pm
ik(\varepsilon)\xi}\psi_\varepsilon^\pm(\xi)$, where $k(\varepsilon)$
is the only positive root of Eq.(\ref{DisRel}) and
$\psi_\varepsilon^\pm(\xi) \to 1$ as $\varepsilon\to 0$. The latter
hypothesis is quite natural, since $v(\xi)=e^{\pm ik(\varepsilon)\xi}$
are solutions of $(L_\varepsilon-F'(0))v=0$.

Then for small $\varepsilon$, the term $H_\varepsilon\tilde u$ in right-hand side
of the inhomogeneous equation (\ref{perturbation}) dominates and the
solvability condition for this inhomogeneous equation \cite{V1}
can be written approximately as the orthogonality condition
\begin{eqnarray}
0 = J_{\pm}(\varepsilon)
&\approx& \int_{-\infty}^\infty e^{\pm
ik(\varepsilon)\xi}H_\varepsilon\tilde u(\xi)~d\xi.\label{OrtCond}
\end{eqnarray}
The asymptotic value of the integral in ({\ref{OrtCond}) as
$\varepsilon\to 0$ is determined by the closest to the real axis
singularities of integrand in the complex plane (the {\it Darboux
principle}, see \cite{Boyd}). Since $H_\varepsilon \tilde u(\xi)$ is
even, bounded, and real-valued for real $\xi$, the main contribution
comes from $z_\pm=\pm\alpha+i\beta$ and $J_+ = J_- \equiv J$.

In the simplest case, when the integrand has poles of order $n$
in the points $z_\pm$, the result is simply a sum of the residues in these poles multiplied by $2\pi i$.
In a more complicated case, the singularities $z_\pm$  of $H_\varepsilon  \tilde u$ can be
rational or transcendental branch points. For both cases, since
$H_{\varepsilon} \tilde{u}(x)$ is even and real for real $x$, we can write
\begin{eqnarray}
\label{sing-1} H_\varepsilon  \tilde u(\xi) \sim
C(\varepsilon)e^{i\pi\kappa/2}(\xi-z_+)^\kappa,\quad \xi\to z_+
\\
H_\varepsilon  \tilde u(\xi) \sim
\overline{C(\varepsilon)}e^{i\pi\kappa/2}(\xi-z_-)^\kappa,\quad \xi\to
z_-,
\end{eqnarray}
where $\kappa$ is a real number, $\kappa\ne 0,1,2,\ldots$. It is
naturally to assume that $C(\varepsilon) \sim C_0 \varepsilon^q$ as
$\varepsilon \to 0$ for some values of $C_0$ and $q$. Then applying
standard formulas \cite{Murray} in the asymptotic limit $k(\varepsilon)
\to \infty$ as $\varepsilon \to 0$, we conclude that
\begin{eqnarray}
J(\varepsilon)
& \sim &\frac{4\pi \varepsilon^q |C_0| e^{-\beta
k(\varepsilon)}}{(k(\varepsilon))^{\kappa+1}\Gamma(-\kappa)}\cos\left(\alpha
k(\varepsilon)+\phi_0 \right),\label{asymptotics}
\end{eqnarray}
where $\phi_0 = {\rm arg}(C_0)$. Consequently, zeros of $J(\varepsilon)$ obey the
asymptotic formula (\ref{AsRel}) with $\varphi_0 = \pi/2 - \phi_0$.

If $\tilde u(\xi)$ is a symmetric kink solution of Eq.(\ref{lim-eq}),
the reasoning remains the same up to the point that $H_\varepsilon
\tilde u(\xi)$ is now odd in $\xi$.

\vspace{0.5cm}

{\bf Examples.} The validity of the main result has been confirmed by
many numerical studies. Below we give three illustrative
examples that concern problems of different physical origins.

{\it Example 1}. Consider the equation
\begin{equation}
  \varepsilon^2u''''+u''-u + ru^2-u^3=0, \label{PertSol}
\end{equation}
where $\varepsilon$ is a parameter.  Eq.(\ref{PertSol}) arises in
hydrodynamics where it describes travelling waves for the fifth-order KdV
equation \cite{Kaw1}. If $\varepsilon=0$ and $r>3/\sqrt{2}$,
Eq.(\ref{PertSol}) has an exact soliton solution
\begin{equation}
  \tilde u(\xi) = \frac{3}{\sqrt{r^2-\frac{9}{2}}\cosh \xi+r}.\label{Soln_e=0}
\end{equation}
Closest to the real axis singularities in the upper complex half-plane
are the simple poles
\begin{equation}
  z_\pm(r)= \pm \arctanh{\frac{3}{\sqrt{2}r}} + i\pi.
  \label{sing_soliton}
\end{equation}
Since $H_{\varepsilon} = -\varepsilon^2 \partial_\xi^4$, we note that
the singularities of $H_{\varepsilon} \tilde u$ are situated in
(\ref{sing_soliton}) and are poles
 of order $n = 5 = -\kappa$. The expansion
(\ref{sing-1}) holds with $C(\varepsilon) = \varepsilon^2 C_0$, where
$C_0$ is purely imaginary. Therefore, $\phi_0 = \pi/2$ in Eq.
(\ref{asymptotics}) and $\varphi_0 = 0$ in Eq. (\ref{AsRel}).

The dispersion relation (\ref{DisRel}) reads as $\varepsilon^2 k^4 - k^2 - 1 = 0$
and it has one positive root $k_0(\varepsilon)$ such that
$k_0(\varepsilon) \sim 1/\varepsilon$ as $\varepsilon\to 0$. According to the main result, we
expect that there exists an infinite sequence of values $\{ \varepsilon_n \}$
such that Eq.(\ref{PertSol}) has soliton solutions for $\varepsilon = \varepsilon_n$
with the asymptotic formula
\begin{equation}
\pi n \varepsilon_n \sim \alpha = \arctanh{\frac{3}{\sqrt{2}r}} \quad \mbox{\rm as} \quad n \to \infty.
\label{Ex1As}
\end{equation}

Numerical computations strongly support this prediction.
Fig.\ref{Ex1Line} shows the values $\alpha/(\pi \varepsilon_n)$
which approach to integers for larger values of $n$. The profiles of the three
lowest solitons corresponding to points $A$, $B$, and $C$ are shown on the inserts by solid line,
together with the limiting soliton (\ref{Soln_e=0}) by dotted line. The discrepancy
reduces quickly for larger values of $n$.

\begin{figure}[h]
\centerline{\includegraphics [scale=0.5]{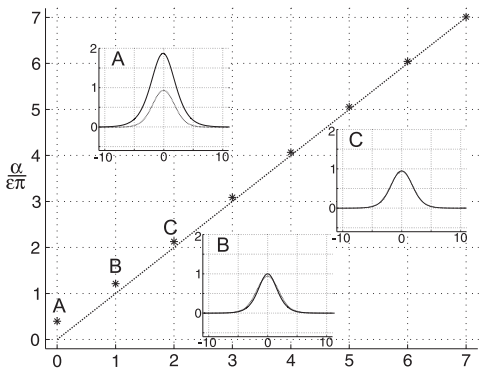}}
\caption{Soliton solutions of Eq.(\ref{PertSol}) with $r=2.3$. Values
of $\alpha/(\pi \varepsilon_n)$ are shown by asterisks. Profiles of the
first three solitons are shown on the inserts by solid line, the dotted
line shows the limiting soliton (\ref{Soln_e=0}). }\label{Ex1Line}
\end{figure}

{\it Example 2}. Consider the nonlocal double sine-Gordon equation
\begin{equation} \label{nonlocalKG}
u_{tt} = \int_{\mathbb{R}}
K_{\varepsilon}(|x-y|) u_{yy} dy + \sin u + 2a \sin 2u,
\end{equation}
where $a > 0$ is a parameter. In particular, this equation arises in
nonlocal Josephson electrodynamics where it describes layered
structures \cite{AOSU95} (the second sine harmonic is important if they
include, for instance, ferromagnetic layers, \cite{GKI04}). A list of
possible kernels $K_\varepsilon$ which arises in Josephson models can
be found in \cite{SUST09}. We assume that $K_0$ is the Dirac
distribution such that Eq. (\ref{nonlocalKG}) with $\varepsilon = 0$
reduces to the classical double sine-Gordon equation. If we denote the
Fourier transform of $K_\varepsilon$ by ${\hat K}_\varepsilon(k)$, then
${\hat K}_\varepsilon(k)\to {\hat K}_0 = 1$ as $\varepsilon\to 0$.

Travelling wave solutions $u(\xi)=u(x-vt)$ of Eq. (\ref{nonlocalKG})
satisfy the equation
\begin{eqnarray}
v^2u_{\xi\xi} = \int_{\mathbb{R}} K_{\varepsilon}(|\xi-\xi'|)
u_{\xi'\xi'} d\xi' + \sin u + 2a \sin 2u. \label{TrW_2SG}
\end{eqnarray}
If $\varepsilon=0$, Eq.(\ref{TrW_2SG}) reads
\begin{eqnarray}
\label{lim-kink}
(1-v^2)u'' = \sin u + 2a\sin 2u,
\end{eqnarray}
where $v^2 < 1$ is assumed. We consider $2\pi$-kink solutions with boundary conditions at infinity
\begin{eqnarray*}
\lim_{\xi\to-\infty}u(\xi)=0, \quad \lim_{\xi\to+\infty}u(\xi)=2\pi.
\end{eqnarray*}
For $a > 0$, Eq. (\ref{lim-kink}) has exact $2\pi$-kink solution
\begin{eqnarray}
\tilde{u}(\xi)=\pi + 2\arctan\left(\frac1{\sqrt{1+4a}}
\sinh\left(\frac{\sqrt{1+4a}}{\sqrt{1-v^2}}\xi\right)\right) \label{2SG_Sol}
\end{eqnarray}
and the closest singularities to the real axis are the two logarithmic branching points
$z_\pm=\pm\alpha+i\beta$, where
\begin{eqnarray*}
\alpha=\frac{\sqrt{1-v^2}}{2 \sqrt{1+4a}} {\rm arccosh}(1 + 8a),\quad
\beta =\frac{\pi \sqrt{1-v^2}}{2\sqrt{1+4a}}.
\end{eqnarray*}

The dispersion relation
\begin{eqnarray*}
-v^2k^2+k^2{\hat K}_\varepsilon(k) = 1 + 4a
\end{eqnarray*}
is assumed to have a single pair of real roots $k = \pm k(\varepsilon)$ for all $v^2 < 1$.
In particular, if $K_{\varepsilon}$ is the Kac-Baker kernel
\begin{eqnarray}
K_\varepsilon(|\zeta|)=\frac1{2\varepsilon}\exp\left(-\frac{|\zeta|}{\varepsilon}\right),\quad
{\hat K}_\varepsilon(k)=\frac1{1+\varepsilon^2k^2},\label{KB}
\end{eqnarray}
then this assumption is satisfied and
\begin{eqnarray*}
k_0(\varepsilon)\sim \frac{\sqrt{1-v^2}}{\varepsilon v}, \quad {\rm as} \quad
\varepsilon\to 0.
\end{eqnarray*}

Let us present arguments that Eq.(\ref{TrW_2SG}) with the kernel
(\ref{KB}) admits an infinite sequence of the $2\pi$-kink solutions.
We note that this equation  can be reduced to the system of
differential equations
\begin{eqnarray}
\left\{ \begin{array}{l}
v^2u_{\xi\xi} = q + \sin u + 2a \sin 2u, \\
-\varepsilon^2q_{\xi\xi} + q = u_{\xi\xi}, \end{array} \right.
\label{Eq2}
\end{eqnarray}
where an additional variable $q$ is introduced. For the limiting kink $\tilde{u}$,
we denote a solution of the second equation of the system (\ref{Eq2}) by $\tilde{q}$.
Now  since $H_{\varepsilon} \tilde{u}  = \tilde{u}'' - \tilde{q}$,
we understand that $H_{\varepsilon} \tilde{u}$ has a double pole in $\tilde{u}''$
at $z_\pm=\pm\alpha+i\beta$ in addition to the logarithmic branching points in $\tilde{q}$.
Hence $n = 2 = -\kappa$ and
the expansion (\ref{sing-1}) holds with $C(\varepsilon) \to C_0$
as $\varepsilon \to 0$, where $C_0$ is real.
Therefore, $\phi_0 = 0$  in Eq. (\ref{asymptotics}) and $\varphi_0 = \pi/2$ in Eq. (\ref{AsRel}).

According to the main result, we
expect that there exists an infinite sequence of values $\{ \varepsilon_n \}$
such that Eq.(\ref{TrW_2SG}) with the kernel (\ref{KB}) has a
$2\pi$-kink solution for $\varepsilon = \varepsilon_n$ with the asymptotic formula
as $n \to \infty$,
\begin{equation}
\pi (1 + 2 n) \varepsilon_n \sim \delta = \frac{(1-v^2)  {\rm arccosh}(1 + 8a)}
   {v \sqrt{1+4a}}.\label{Ex2As}
\end{equation}

Using an appropriate shooting method \cite{AlfMed11}, we compute numerically
the values of $\varepsilon$, for which there exist $2\pi$-kink solutions of
Eq. (\ref{TrW_2SG}) with the kernel (\ref{KB}).
Numerical calculations strongly confirm the existence of the sequence
$\{ \varepsilon_n \}$ as well as its asymptotic properties (\ref{Ex2As}). The
values of $\delta/(\pi \varepsilon_n)$ for $a = 1/8$ and $v = 0.1$
are given in Table \ref{Table1} and approach closer to odd integers
for larger values of $n$.

Evidently, each value $\varepsilon_n$ depends on the parameter $v$;
however from the physical viewpoint, the inverse functions
$v_n(\varepsilon)$ are more important. Fig.\ref{fig-example-2}
represents the dependence of the velocities $v_n$ versus $\varepsilon$
for the first three $2\pi$-kink solutions. The corresponding profiles
of the $2\pi$-kinks (solid line)  at the points $A$, $B$ and $C$ are
shown in the inserts together with the limiting kink (\ref{2SG_Sol})
(dotted line). The difference between the actual kink and the limiting
kink (\ref{2SG_Sol}) is not visible already for kinks at the points $B$
and $C$.

\begin{table}[h]
\begin{tabular}{|l|c|c|c|c|c|c|}  \hline
$1 + 2 n$ & 1 & 3 & 5 & 7 & 9 & 11  \\ \hline $\delta/(\pi
\varepsilon_n)$ & 3.7168 & 4.9763 & 6.3699 & 7.8595 & 9.4541 & 11.1396
\\ \hline
\end{tabular}
\caption{The values of $\delta/(\pi \varepsilon_n)$ for which Eq.(\ref{TrW_2SG}) with the kernel (\ref{KB})
admits $2\pi$-kink solution for $a=1/8$ and $v=0.1$.}\label{Table1}
\end{table}

\begin{figure}[h]
\centerline{\includegraphics[scale=0.5]{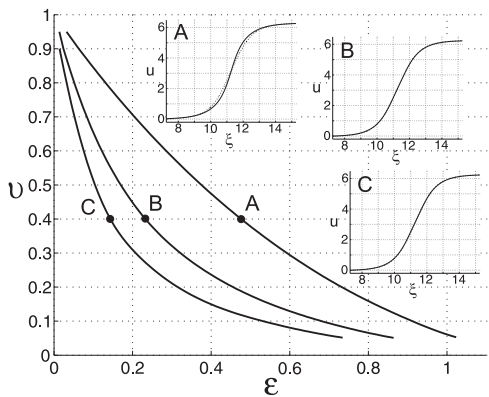}}
 \caption{Kink solutions of Eq.(\ref{TrW_2SG}) with the kernel (\ref{KB}).
Values of $v_n$ versus $\varepsilon$ are shown for the first three
solutions. Profiles of the first three kinks are shown in the inserts
by solid line, the dotted line shows the limiting kink
(\ref{2SG_Sol}).} \label{fig-example-2}
\end{figure}

{\it Example 3}. Discrete Klein-Gordon equation is one of the basic equations describing lattice dynamics in various
contexts, from solid state physics to biophysics \cite{Savin}. Travelling
waves of the discrete Klein--Gordon equation satisfy the equation
\begin{equation}
\label{adv-del_TrW}  v^2u_{\xi\xi} = \varepsilon^{-2}(u(\xi+\varepsilon)-2u(\xi)+u(\xi-\varepsilon))
+ F(u),
\end{equation}
where $\varepsilon$ is the spacing between lattice sites and $F$ is a nonlinear
function. Bistable nonlinearity
\begin{eqnarray}
F(u)=u (1-u^2)(1+\gamma u^2), \quad \gamma > 0,
\label{fi4-6}
\end{eqnarray}
may support kinks which satisfy the boundary conditions
\begin{eqnarray*}
\lim_{\xi\to\pm\infty}u(\xi)=\pm 1.
\end{eqnarray*}
If $\gamma = 0$, Eq.(\ref{adv-del_TrW})  corresponds to the classical
$\phi^4$ model, where no travelling kinks were previously found
\cite{IP}. We anticipate that for $\gamma > 0$ and $v$ fixed there
exists an infinite sequence of travelling kinks for discrete values of
parameter $\varepsilon$.

If $\varepsilon=0$, Eq.(\ref{adv-del_TrW}) reads
\begin{eqnarray}
\label{lim-kink-phi}
(1-v^2) u'' + u (1-u^2)(1+\gamma u^2) = 0,
\end{eqnarray}
where $v^2 < 1$ is assumed. For $\gamma > 0$, Eq. (\ref{lim-kink-phi}) has exact kink solution
\begin{eqnarray}
\tilde{u}(\xi)= \frac{\sqrt{3+\gamma} \tanh(\eta \xi)}{\sqrt{3(1+\gamma)-2 \gamma \tanh^2(\eta \xi)}},
\; \eta = \frac{\sqrt{1+\gamma}}{\sqrt{2(1-v^2)}}
\label{2SG_Sol-phi}
\end{eqnarray}
and the closest singularities to the real axis are the two square root branching points
$z_\pm=\pm\alpha+i\beta$, where
\begin{eqnarray*}
\alpha=\frac{\sqrt{1-v^2}}{\sqrt{2(1+\gamma)}} {\rm arccosh}\left(\frac{3 + 5 \gamma}{3 + \gamma}\right),\quad
\beta =\frac{\pi \sqrt{1-v^2}}{\sqrt{2(1+\gamma)}}.
\end{eqnarray*}

If $\gamma>0$ the dispersion relation
\begin{eqnarray*}
-v^2k^2+ 4 \varepsilon^{-2} \sin^2(k \varepsilon/2) + 2(1+\gamma) = 0
\end{eqnarray*}
has a single pair of real roots $k = \pm k(\varepsilon)$ for $v \in
(v_0,1)$ for some $v_0\equiv v_0(\gamma)>0$. More than one pair of
roots exist (always, an odd number) for $v \in (0,v_0)$. We note that
$k(\varepsilon) \sim p_0/\varepsilon$ as $\varepsilon \to 0$, where
$p_0$ is a positive root of the transcendental equation $-v^2 p_0^2 + 4
\sin^2(p_0/2) = 0$. This equation has a single pair of real roots for
$v \in (\tilde v_0,1)$ where $\tilde v_0 \approx 0.22$ \cite{IP}.

For the limiting kink $\tilde{u}$, we have
$$
H_{\varepsilon} \tilde{u}  = \tilde{u}'' - \varepsilon^{-2}(\tilde{u}(\xi+\varepsilon)-2 \tilde{u}(\xi) +
\tilde{u}(\xi-\varepsilon)).
$$
The first term yields the expansion (\ref{sing-1}) with $\kappa = -5/2$ and
$C(\varepsilon) \sim C_0 e^{5 i \pi/4}$ as $\varepsilon \to 0$
near the singular points $z_\pm =\pm\alpha+i\beta$, where $C_0$ is purely imaginary.
The second term yields the expansion (\ref{sing-1}) with $\kappa = -1/2$ and
$C(\varepsilon) \sim \varepsilon^{-2} C_0 e^{i \pi/4}$ as $\varepsilon \to 0$, where $C_0$ is purely imaginary.
Both terms yield equal contribution in $\varepsilon$
that depends on $v$. Nevertheless, for both terms,
$\phi_0 = -\pi/4$ in Eq. (\ref{asymptotics}) and
$\varphi_0 = 3 \pi/4$ in Eq. (\ref{AsRel}).

According to the main result, we
expect that there exists an infinite sequence of values $\{ \varepsilon_n \}$
such that Eq.(\ref{adv-del_TrW}) has a kink solution for $\varepsilon_n$
with the asymptotic formula
\begin{equation}
\pi (4 n + 3) \varepsilon_n \sim \chi = 4 p_0 \alpha
\quad \mbox{\rm as} \quad n \to \infty.    \label{Ex3As}
\end{equation}

Using Newton's method with the fourth-order finite-difference approximation of
the second derivative, we compute numerically the values of $\varepsilon$,
for which kink solutions of  Eq.(\ref{adv-del_TrW}) exist.
Numerical calculations strongly confirm the existence of the sequence
$\{ \varepsilon_n \}$ as well as its asymptotic properties (\ref{Ex3As}). The
values of $\chi/(\pi \varepsilon_n)$ for $\gamma = 5$ and $v = 0.6$
are given in Table \ref{Table2} with satisfactory agreement.
\begin{table}[h]
\begin{tabular}{|l|c|c|c|c|}  \hline
$3 + 4 n$ & 3 & 7 & 11 & 15  \\ \hline
$\chi/(\pi \varepsilon_n)$ & 3.5303 & 7.3547 & 11.1520 & 15.0329 \\ \hline
\end{tabular}
\caption{The values of $\varepsilon$ for which Eq.(\ref{adv-del_TrW})
admits kink solutions for $\gamma = 5$ and $v = 0.6$.}\label{Table2}
\end{table}

Fig.\ref{fig-example-3} shows the first two kink solutions of
Eq.(\ref{adv-del_TrW}) with $\gamma = 5$ and $v = 0.6$ (thick and thin
solid lines) together with the limiting solution of
Eq.(\ref{2SG_Sol-phi}) (dotted line). The difference between the third
kink solution and the limiting kink is not visible in the figure.

\begin{figure}[h]
\centerline{\includegraphics[scale=0.4]{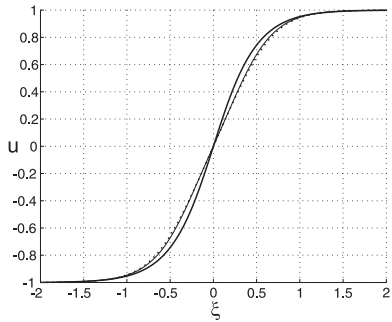}}
 \caption{Profiles of the first (thick solid line) and second (thin solid line) kink solutions
 and the limiting solution (\ref{lim-kink-phi}) (dotted line) for Eq.(\ref{adv-del_TrW}) with
 $\gamma = 5$ and $v = 0.6$.}
\label{fig-example-3}
\end{figure}

\vspace{0.5cm}

{\bf Conclusion.} We have shown on three prototypical examples that
Hamiltonian systems with resonances may admit
an infinite sequence of travelling solitons or kinks
if the leading-order asymptotic solution has a pair of
symmetric singularities in the upper half-plane.
This simple but universal observation reveals the way why travelling
solitons or kinks have increased mobility in some nonlinear systems with
resonances but not in the others.

To emphasize the universality of this prediction, we mention two
additional examples earlier studied in literature, where the mechanism
for existence of the infinite sequence of travelling solitons or kinks
also should take place. First, three families of radiationless
travelling solitons were captured numerically in the saturable discrete
nonlinear Schr\"{o}dinger equation \cite{Mel1}. Although the
leading-order asymptotic solution is not available in the closed form,
one can show that it has a pair of symmetric singularities in the upper
half-plane, hence one can anticipate that there exist infinitely many
families of travelling solitons in this example. Second, two moving
kinks were reported in the modified Peyrard--Remoissenet potential of
the discrete Klein--Gordon equation \cite{Savin2000}. Since this model
is a modification of the discrete double sine--Gordon equation, one can
anticipate the existence of an infinite sequence of families of the
moving kinks in this example. Our study of some other models of this
kind (e.g. various nonlocal generalizations of double and triple
sine-Gordon model) also confirms the asymptotical formula
(\ref{AsRel}).

\textbf{{Acknowledgments.}} Authors are grateful to A.S.Malishevskii
for useful discussions. The work of GLA and EVM was supported by
Russian federal program ``Scientific and educational personnel of the
innovative Russia'', grant 14.B37.21.1273. The work of DP is supported
by the ministry of education and science of Russian Federation (Project
14.B37.21.0868).


\begin{thebibliography}{99}

\bibitem{Kaw1} T. Kawahara, 
J. Phys. Soc. Japan {\bf 33}, 260 (1972);
J.K. Hunter and J. Scheule, 
Physica D {\bf 32}, 253 (1988); A.R. Champneys and M.D. Groves,
     J. Fluid Mech. {\bf 342}, 199 (1997); A.R. Champneys, 
Nonlinearity {\bf 14}, 87 (2001);
A. Tovbis and D. Pelinovsky, 
Nonlinearity  {\bf 19}, 2277 (2006)

\bibitem{Savin} Y. Zoloratyuk, J.C. Eilbeck, and A.V. Savin,
Physica D {\bf 108}, 81 (1997);
P.G. Kevrekidis, 
Physica D {\bf 183}, 68 (2003);
A.A. Aigner, A.R. Champneys, and V.M. Rothos, 
Physica D {\bf 186}, 148 (2003)


\bibitem{Savin2000}A.V. Savin, Y. Zolotaryuk, and J.C. Eilbeck, Physica D {\bf 138}, 267
(2000);


\bibitem{IP} G. Iooss and D.E. Pelinovsky, 
Physica D {\bf 216}, 327 (2006);
O.F. Oxtoby, D.E. Pelinovsky, and I.V. Barashenkov, 
Nonlinearity {\bf 19}, 217 (2006)


\bibitem{A1} G.L. Alfimov, V.M. Eleonsky, and N.E. Kulagin, 
Chaos {\bf 2}, 565 (1992); G.L. Alfimov, V.M. Eleonsky, N.E. Kulagin, and N.V.Mitzkevich,
Chaos {\bf 3}, 405 (1993); G.L. Alfimov, V.M. Eleonsky, and L.M. Lerman,
Chaos {\bf 8}, 257 (1998)

\bibitem{AOSU95} Yu. M. Aliev, K.N. Ovchinnikov, V.P. Silin, S.A. Uryupin,
JETP {\bf 80}, 551 (1995); S. Savel'ev, V. Yampol'skii, A. Rakhmanov, F. Nori,
  Rep. Prog. Phys. {\bf 73}, 026501 (2010)

\bibitem{ES3Gen}
 A.R. Champneys, B.A. Malomed, and M.J. Friedman, 
     Phys. Rev. Lett. {\bf 80}, 4168 (1998);
     A.R. Champneys, B.A. Malomed, J. Yang, and D.J. Kaup, 
     Physica D {\bf 152-153}, 340 (2001);
T. Wagenknecht and A.R. Champneys, 
     Physica D {\bf 177}, 50 (2003)

\bibitem{G1} V.G. Gelfreich, V.F. Lazutkin, and M.B. Tabanov, Chaos {\bf 1}, 137 (1991);
V. Gelfreich and C. Simo, Discr. Cont. Dynam. Syst. {\bf B} {\bf 10},
511 (2008).

\bibitem{V1} V. Vougalter and V. Volpert, 
Proc. Edinb. Math. Soc. {\bf 54}, 249 (2011);
Anal. Math. Phys. {\bf 2012}, 473 (2012).

\bibitem{Boyd} J. Boyd, {\em  Chebyshev and Fourier Spectral Methods} (Dover Publishers, 2001)


\bibitem{GKI04} A. A. Golubov, M. Yu. Kupriyanov, and E. Il'ichev, Review of
Modern Physics {\bf 76}, 411 (2004);


\bibitem{SUST09}
    A.A. Abdumalikov, G.L. Alfimov, A.S. Malishevskii,
    Supercond. Sci. Technol. {\bf 22}, 023001 (2009)













\bibitem{Murray} J.D. Murray, {\it Asymptotic Analysis} (Springer, 1984)






\bibitem{AlfMed11}
    G.L. Alfimov and E.V. Medvedeva, 
    Phys. Rev. E. {\bf 84}, 056606 (2011)


\bibitem{Mel1} T.R.O. Melvin, A.R. Champneys, P.G. Kevrekidis, and J. Cuevas, Phys. Rev. Lett. {\bf 97},
124101 (2006); Physica D {\bf 237}, 551 (2008).


\end{thebibliography}
\end{document}